\documentstyle[12pt]{article}
\newcommand{\beq}{\begin{equation}}
\newcommand{\eeq}{\end{equation}}
\newcommand{\ba}{\begin{eqnarray}}
\newcommand{\ea}{\end{eqnarray}}

\newcommand{\dsl}
  {\kern.06em\hbox{\raise.15ex\hbox{$/$}\kern-.56em\hbox{$\partial$}}}

\newcommand{\eeqarr}{\end{eqnarray}}
\newcommand{\ZZ}{{\rm \kern 0.275em Z \kern -0.92em Z}\;}
\begin{document}
\begin{titlepage}
\begin{center}
{\Huge Multi-Trace Operators and the Generalized}
\\
{\Huge AdS/CFT Prescription}
\\
\vspace*{2.0cm}
{\large Pablo Minces}
\\
\vspace*{0.5cm}
Instituto de F\'{\i}sica Te\'orica, Universidade Estadual
Paulista\\
Rua Pamplona 145, 01405-900, S\~ao Paulo, SP, Brasil\\
pablo@fma.if.usp.br
\vspace*{1.5cm}

\end{center}
\begin{abstract}
We show that multi-trace interactions can be consistently incorporated
into an extended AdS/CFT prescription involving the inclusion of
generalized boundary conditions and a modified Legendre transform
prescription. We find new and consistent results by considering a
self-contained formulation which relates the quantization of the bulk
theory to the AdS/CFT correspondence and the perturbation at the boundary
by double-trace interactions. We show that there exist particular
double-trace perturbations for which irregular modes are allowed to
propagate as well as the regular ones. We perform a detailed analysis of
many different possible situations, for both minimally and non-minimally
coupled cases. In all situations, we make use of a new constraint which is
found by requiring consistency. In the particular non-minimally coupled
case, the natural extension of the Gibbons-Hawking surface term is
generated.

\end{abstract}

\vskip 3.0cm

\begin{flushleft}
\end{flushleft}
\end{titlepage}

\section{Introduction}

In recent papers \cite{berkooz1}\cite{berkooz3}, the interest on the role
of multi-trace interactions in the AdS/CFT correspondence \cite{maldacena}
has been revived by introducing the proposal that, to deform
the boundary CFT by double-trace operators, gives rise to a new
perturbation expansion for string theory which is based on a non-local 
worldsheet. 

In the AdS/CFT picture, multi-trace operators in
the boundary CFT are understood as multi-particle states in AdS
(see, for example, \cite{mald}\cite{freedman9}), and this raises the
question of how to place
a boundary condition on a multi-particle state and perform explicit
calculations in the context of the AdS/CFT correspondence. 

A solution to this problem came with the proposal in
\cite{witten7}\cite{berkooz2} that multi-trace
operators can be incorporated by
generalizing the usual Dirichlet prescription which is considered in the
case of single-trace operators. Such prescription reads 
\cite{witten}\cite{gubser}

\beq
\exp\left( -I _{AdS}[\phi_{0}]\right) \equiv \left<\exp\left(\int d^{d}x
\;
{\cal O}(\vec{x}) \; \phi_{0}(\vec{x})\right)\right>,
\label{10}
\eeq
where $d+1$ is the dimension of the AdS bulk, and $\phi_{0}$ is the
boundary value of the bulk field $\phi$ which
couples to the boundary CFT operator ${\cal O}$.

The perturbation of the boundary CFT by
multi-trace operators can be written schematically as

\beq
I_{QFT}[{\cal O}]=I_{CFT}[{\cal O}] + \int d^{d}x\; W[{\cal O}]\; ,
\label{100}
\eeq
where $W[{\cal O}]$ is an arbitrary function of ${\cal O}$. In the
particular case of a double-trace perturbation of the form

\beq
W[{\cal O}]=\frac{\beta}{2}\;{\cal O}^{2}\; ,
\label{101}
\eeq
where ${\cal O}$ has conformal dimension $\Delta=d/2$, it has been shown
in \cite{witten7} that a generalized boundary condition gives rise to
the correct renormalization formula for the coupling $\beta$.

The purpose of this paper is to introduce a further development of the
proposal in \cite{witten7}\cite{berkooz2} by considering a
generalized
AdS/CFT prescription in which multi-trace interactions can be consistently
incorporated. More specifically, we aim at showing that the recent
proposal in \cite{our3} for a generalized AdS/CFT prescription which
makes use of generalized boundary conditions provides a natural
frame for consistently describing multi-trace operators in the AdS/CFT
correspondence. For the sake of simplicity, we will concentrate on
the particular case of double-trace
interactions, for which we will relate the coupling coefficient to
specific boundary conditions on the scalar field, or equivalently, to the
addition of specific boundary terms to the usual bulk action. We will also
interpret our results in terms of the energy of the
theory in the bulk and the constraints which arise when performing its
quantization. In addition, we will extend the formulation to the
non-minimally coupled case.

In considering the quantization of the scalar field theory on AdS, it
has been shown in \cite{freedman}\cite{freedman8}\cite{mezincescu} that 
there exist particular constraints on the mass and the coupling
coefficient to the metric for which two different asymptotic behaviors of
the field make the energy to be conserved, positive and finite. Such
asymptotic behaviors are of the form

\beq
\phi^{R}\sim\epsilon^{\Delta_{+}}\; ,\qquad\quad
\phi^{I}\sim\epsilon^{\Delta_{-}}\; ,
\label{108}
\eeq 
where $\phi^{R}$ and $\phi^{I}$ stand for `regular' and
`irregular' modes, $\epsilon$ is a measure of the distance to the boundary
which is considered to be small, and

\beq
\Delta_{\pm}= \frac{d}{2}\pm\nu\; ,
\label{1000}
\eeq
where

\beq
\nu=\sqrt{\frac{d^2}{4}\;+\;M^{2}}\; .
\label{11}
\eeq
Here $M$ is the effective mass of the scalar field, and in the
non-minimally coupled case it is given by

\beq
M^{2}=m^{2}+\varrho R\; ,
\label{11''}
\eeq
where $R$ is the Ricci scalar of the metric (which is a
negative constant) and $\varrho$ is the coupling coefficient. In the
minimally coupled case $\varrho =0$, the equation above reduces to $M=m$.

Since there exist two possible quantizations of the theory on the bulk,
we expect that there must exist two different boundary CFT's. However, it
has been pointed out in \cite{witten2} that the usual Dirichlet
prescription Eq.(\ref{10}) can only account for one of them, namely, the
one with conformal dimension $\Delta_{+}$, which corresponds to regular
modes propagating in the bulk. In order to also account for the missing
conformal dimension $\Delta_{-}$, the proposal in \cite{witten2} is that
its generating functional is the Legendre transform of the one which gives
rise to the conformal dimension $\Delta_{+}$. It was explicitly shown
in \cite{witten2} that in fact this prescription gives rise to the missing
conformal dimension $\Delta_{-}$.

Later, it was shown in \cite{our3} that there were still some problems
which needed to be considered. One of them is the fact that the usual
AdS/CFT prescription is unable to reproduce the constraints in $\varrho$
and $\nu$
for which the energy in \cite{freedman}\cite{freedman8}\cite{mezincescu}
is conserved, positive and finite for irregular modes propagating in the
bulk. The second problem is that the usual Legendre transform
prescription leaves
a coefficient to be fixed `by hand'.  And the last problem regarding the
usual Legendre transform is that it does not work for some particular
values of $\nu$.

Then, the proposal in \cite{our3} was to consider a generalized formalism
which
involves modifications both in the bulk quantization and in the AdS/CFT
prescription. From the bulk point of view, it was suggested in \cite{our3}
that the natural energy to be considered in the AdS/CFT correspondence
context is the `canonical' energy which is constructed out of the Noether
current corresponding to time translations, rather than the `metrical' one
which is constructed through the stress-energy tensor, as in
\cite{freedman}\cite{freedman8}\cite{mezincescu}. The reason for
considering the canonical energy is that, unlike the metrical one, it is
sensitive to the addition of boundary terms to the action, as it
happens to the AdS/CFT prescription Eq.(\ref{10}).

Finally, from the AdS/CFT correspondence point of view, the proposal in
\cite{our3} was
to consider a generalized AdS/CFT prescription of the form (see also
\cite{our2} for previous results)
 
\beq
\exp\left( -I _{AdS}[f_{0}]\right) \equiv \left<\exp\left(\int d^{d}x \;
{\cal O}(\vec{x}) \; f_{0}(\vec{x})\right)\right>,
\label{13}
\eeq
where, unlike the usual prescription Eq.(\ref{10}), we make use of a
generalized source $f_{0}$ which depends on the boundary 
conditions.\footnote{ Note that the addition of boundary terms to the
action changes
the boundary conditions under which the action is stationary, and on the
other hand it also changes the generating functional for the boundary
CFT. This is the reason why considering generalized boundary conditions
as in Eq.(\ref{13}) involves more information than making use of the 
usual prescription Eq.(\ref{10}).} The formulation in \cite{our3} makes
use of Dirichlet, Neumann and mixed boundary conditions on the scalar
field in both the minimally and non-minimally coupled cases. In addition, 
it involves a generalized Legendre transform
prescription of the form

\beq
{\tilde I}_{AdS}[f_{0},{\tilde f}_{0}]\; = I_{AdS}[f_{0}]\; -\;
\int d^{d}x\;
f_{0}\left(\vec{x}\right){\tilde f}_{0}\left(\vec{x}\right)\; ,
\label{14}
\eeq
which differs from the usual one in \cite{witten2} by the fact that it
involves the whole on-shell action rather than only the leading non-local
term. It was shown in \cite{our3} that the divergent local terms of the
action contain information about the transformed generating functional,
and then they have to be taken into account. In addition, the prescription
above leaves no coefficient to be fixed `by hand'. After performing
the Legendre transformation, the transformed boundary CFT can be
determined from the prescription

\beq
\exp\left( -{\tilde I}_{AdS}[{\tilde f}_{0}]\right) \equiv
\left<\exp\left(\int
d^{d}x \;
{\tilde {\cal O}}(\vec{x}) \; {\tilde f}_{0}(\vec{x})\right)\right>\; .
\label{15}
\eeq

It was shown in \cite{our3} that this generalized Legendre
transform prescription removes all problems mentioned above regarding the
usual prescription. In addition, the key result in
\cite{our3} has been to show that the generalized prescription
Eqs.(\ref{13}-\ref{15}) gives rise to precisely the same constraints
on $\varrho$ and $\nu$ for which irregular modes propagate in the bulk
when the quantization is
performed by making use of the canonical energy rather than the metrical
one.

One of the purposes of this paper is to show that the introduction of
double-trace perturbations at the boundary CFT can be understood in terms
of the formulation in \cite{our3}. This would be a powerful result for
many reasons. The first one is that it would lead in a natural way to 
a generalized AdS/CFT prescription in which multi-trace interactions can 
be consistently incorporated. 

The second reason is that it would enable us to relate the coupling
coefficient
of the double-trace perturbation to specific boundary conditions on the
bulk field, or equivalently, to the addition of specific boundary terms to
the action. In particular, we will show that there exist particular
couplings which
require to consider new boundary conditions that have not been analyzed in
\cite{our3}. In this article, we will also perform a detailed analysis 
of such new boundary conditions. In all cases, we will find the explicit
form of the coupling coefficients and the surface terms.

The third reason is that to consider double-trace operators in terms of
the formulation in \cite{our3} let us to relate the
corresponding coupling coefficients to the constraints on
$\varrho$ and $\nu$ for which irregular modes propagate
in the bulk when the canonical energy is employed instead of the metrical 
one. In particular, we will show that, when such constraints are
satisfied, the couplings of the double-trace perturbations diverge. This 
result is   
consistent with the statement in \cite{witten7} that, as the coupling
grows, the system approaches the condition that is suitable for
quantization to get a field of dimension $\Delta_{-}$. It is also 
important to point out that, in this formulation, the canonical energy in
the bulk depends on the boundary multi-trace perturbations. 

And the last reason why it is useful to analyze double-trace
operators in terms of the formulation in \cite{our3} is that it let us to
extend, in a natural way, the formulation of double-trace perturbations to
the non-minimally coupled case. In particular, an interesting result that
we will find is that the introduction of a double-trace perturbation at
the boundary generates the natural extension of the Gibbons-Hawking
surface term \cite{gibbons} which is added to the Einstein-Hilbert action
in order to have a well-defined variational principle under variations of
the metric.

We also aim at connecting to the recently
proposed formulation in \cite{muck} for an improved correspondence formula
which gives correct boundary field theory correlators for multi-trace
perturbations. In particular, the key observation in
\cite{witten7}\cite{muck} is that
the conformal operator ${\cal O}$ is to be identified with the Legendre
transformed field ${\tilde\phi}_{0}$. In this paper, we claim that, in
order to map to the boundary all the information contained in the bulk, we
also need to
identify, in the transformed formulation, the conjugated operator
${\tilde{\cal O}}$ with the field $\phi_{0}$. In particular, we will show
that to require consistency imposes a precise 
constraint between the couplings of the double-trace perturbations
corresponding to the conformal operators ${\cal O}$ and ${\tilde{\cal
O}}$. Another difference with respect to the formulation in
\cite{muck} is
that, in this work, we consider that the coupling of the double-trace
perturbation depends on the distance to the boundary. We will show that
this is needed for consistency. In particular,
a result that we will find is that when the constraints for which
irregular modes are allowed to propagate in the bulk are satisfied, the
couplings of
the double-trace perturbations corresponding to the operators ${\cal O}$
and ${\tilde{\cal O}}$ have different asymptotic behaviors. A final
thing to be mentioned is that, in this paper, we will consider the full
functionals containing all local and non-local terms, unlike
the formalism in \cite{muck} which considers only the leading
non-local
term. We make so because the divergent local terms contain information
about the transformed generating functional, and then they have to be
taken into account \cite{our3}.

The paper is organized as follows. In section 2, we revisit the
formulation in \cite{muck} and claim that, in   
order to map to the boundary all the information contained in the bulk, we
also need to
identify, in the transformed formulation, the conjugated operator   
${\tilde{\cal O}}$ with the field $\phi_{0}$. We find the explicit form
of the generating functionals for the boundary CFT's 
in terms of the couplings of the double-trace perturbations
corresponding to the operators ${\cal O}$ and ${\tilde{\cal O}}$. We also
show
that the
requirement for consistency imposes a constraint between such
couplings. In section 3, we analyze the precise way in which the
introduction of double-trace perturbations changes the boundary conditions
on the field by adding surface terms to the action, and extend the
formulation to the non-minimally coupled case. In this situation, we
will show that the introduction of double-trace perturbations generates
the natural extension of the Gibbons-Hawking term. We
also show how double-trace perturbations can be understood in terms of the
generalized AdS/CFT prescription and the new quantization in the bulk as
developed in \cite{our3}. In addition, an interesting result that we
will find is that irregular modes are allowed to propagate precisely when
the
asymptotic behaviors of the couplings of the double-trace perturbations
corresponding to the operators ${\cal O}$ and ${\tilde{\cal O}}$ are
different. In this situation, the coupling corresponding to the
operator of conformal dimension $\Delta_{+}$ diverges. We also
show that a complete formulation requires the inclusion of new
boundary conditions which have not been considered in
\cite{our3}. We perform a detailed analysis of such new boundary
conditions. Finally, section 4 contains our conclusions.

\section{Double-Trace Operators and AdS/CFT}

Throughout this paper, we work in the Euclidean representation of the 
$AdS _{d+1}$ in Poincar\'e coordinates, described by the
half space $x _{0}>0$, $x _{i} \in {\bf R}$ with metric

\beq
ds^{2}=\frac{1}{x _{0}^{2}} \sum_{\mu=0}^{d} dx^{\mu}dx^{\mu},
\label{5}
\eeq
where we have fixed the radius of $AdS_{d+1}$ equal to one. We consider
the space as foliated by a
family of surfaces $x_{0}=\epsilon$ homeomorphic to the boundary at
$x_{0}=0$. The corresponding outward pointing unit normal vector is

\beq
n_{\mu} = \left(-\epsilon^{-1},{\bf 0}\right)\; .
\label{6}
\eeq
The limit $\epsilon\rightarrow 0$ is to be taken only at the end of
calculations \cite{freedman3}.

The usual action for the massive scalar field reads

\beq
I_{0}=\frac{1}{2}\;\int d^{d+1} x \;\sqrt{g}\;
\left(
g^{\mu\nu}\partial_{\mu}\phi\;\partial_{\nu}\phi\; +
\;M^{2}\phi^{2}\right) \; .
\label{1}
\eeq
Performing an infinitesimal variation of the scalar field
$\phi\rightarrow\phi +\delta\phi$, the action above transforms as

\beq
\delta I_{0}=\int
d^{d}x\;\sqrt{h}\;\partial_{n}\phi_{\epsilon}\;\delta\phi_{\epsilon}\; ,
\label{2}
\eeq
where $h_{\mu\nu}$ is the induced metric at the surface
$x_{0}=\epsilon$, $\phi_{\epsilon}$ is the value of the field at
$x_{0}=\epsilon$ and $\partial_{n}\phi$ is the Lie derivative of $\phi$
along $n_{\mu}$. It is given by

\beq
\partial_{n}\phi = n^{\mu}\partial_{\mu}\phi\; .
\label{3}
\eeq
Note that, in Eq.(\ref{2}), the absence of a bulk contribution is due to
the equation of motion

\beq
\nabla^{2}\phi - M^{2}\phi =0\; .
\label{4}
\eeq
The variation Eq.(\ref{2}) shows that the action
$I_{0}$ is stationary for a Dirichlet boundary condition which
fixes
the value of the scalar field $\phi$ at $x_{0}=\epsilon$, namely

\beq
\delta\phi_{\epsilon}=0\; .
\label{7}
\eeq
Integrating by parts and making use of the equation of motion, the action
Eq.(\ref{1}) can be written as the following pure-surface
term

\beq
I_{0} = \frac{1}{2}\int
d^{d}x\;\sqrt{h}\;\phi_{\epsilon}\;\partial_{n}
\phi_{\epsilon}\; .
\label{8}
\eeq

The next step involves to solve the equation of motion and write
$\partial_{n}\phi_{\epsilon}$ in terms of the boundary data
$\phi_{\epsilon}$. This procedure has been carried out in 
\cite{freedman3}\cite{viswa1}, where the final result was expanded in
powers of the distance to the boundary in order to select the
leading non-local term, which is understood as the generating
functional for the boundary CFT \cite{witten}. However, the divergent
local terms of the
action contain information about the Legendre transformed generating
functional \cite{our3}, and then we need to take them into account. This
means that, instead of the leading non-local term, we will make use of the
full action containing all local and non-local terms. It reads

\beq
I_{0}\left[f_{\epsilon}\right] = -\frac{1}{2}\int d^{d}x
\; d^{d}y\;\sqrt{h}\;f_{\epsilon}(\vec{x})
\;f_{\epsilon}(\vec{y})\int\frac{d^{d}k}{\left(
2\pi\right)^{d}}\;e^{-i\vec{k}\cdot\left(
\vec{x}-\vec{y}\right)}\;F(k\epsilon)\; ,
\label{9}
\eeq
where $f_{\epsilon}(\vec{x})$ is the source which couples to the
boundary conformal operator through the prescription
Eq.(\ref{10}), $\vec{x}=(x^{1},...,x^{d})$, $k = \mid\vec{k}\mid$ and

\beq
F(k\epsilon)=\frac{d}{2} + \nu -
k\epsilon\;\frac{K_{\nu+1}(k\epsilon)}{K_{\nu}(k\epsilon)}\; .
\label{12}
\eeq
Here $K_{\nu}$ is the modified Bessel function.

The action Eq.(\ref{9}) is only one of the two functionals which contain
the information about the boundary CFT's. The another one is obtained by
performing the Legendre transformation Eq.(\ref{14}), which gives rise to 
the
following transformed functional containing all local and
non-local terms \cite{our3}

\beq
{\tilde I}_{0}\left[{\tilde f}_{\epsilon}\right]=\frac{1}{2}\int
d^{d}x
\; d^{d}y\;\sqrt{h}\;{\tilde f}_{\epsilon}(\vec{x})
\;{\tilde f}_{\epsilon}(\vec{y})\int\frac{d^{d}k}{\left(
2\pi\right)^{d}}\;e^{-i\vec{k}\cdot\left(
\vec{x}-\vec{y}\right)}\;\frac{1}{F(k\epsilon)}\; .
\label{16}
\eeq
Both functionals $I_{0}$ and ${\tilde I}_{0}$ are needed in order to map
to the boundary all the information contained in the bulk. Each one of the
sources $f_{\epsilon}$ and ${\tilde f}_{\epsilon}$ will couple, after
performing the limit $\epsilon\rightarrow 0$ in a proper way and through
the prescriptions Eqs.(\ref{13}, \ref{15}), to the corresponding
boundary conformal operator.

The important result in \cite{witten7}\cite{muck} is that the transformed
source
${\tilde f}_{0}$ can be identified with the conformal operator
${\cal O}$. In order to do this, we transform back the functional
${\tilde I}_{0}\left[{\tilde f}_{\epsilon}\right]$ and find

\beq
I_{0} = \int d^{d}x
\; d^{d}y\;\sqrt{h}\int\frac{d^{d}k}{\left(
2\pi\right)^{d}}\;e^{-i\vec{k}\cdot\left(
\vec{x}-\vec{y}\right)}\left[\frac{1}{2}\;\frac{1}{F(k\epsilon)}\;{\tilde
f}_{\epsilon}(\vec{x})
\;{\tilde f}_{\epsilon}(\vec{y})\;+
f_{\epsilon}(\vec{x})
\;{\tilde f}_{\epsilon}(\vec{y})\right]\; .
\label{17}
\eeq
Now we will analyze the way in which the results above are affected by
perturbing at the boundary with a double-trace operator of the form

\beq
W[{\cal O}]=\frac{\beta}{2}\; {\cal O}^{2}\; ,
\label{18}
\eeq
where $\beta$ is the coupling coefficient. Following
\cite{witten7}\cite{muck}, the next step is to identify ${\cal
O}$ with ${\tilde f}_{0}$. This would give

\beq
W[{\tilde f}_{0}]=\frac{\beta}{2}\; {\tilde f}_{0}^{2}\; .
\label{18'}
\eeq
However, it is important to note that
the limit $\epsilon\rightarrow 0$ is to be taken only at the very end of
calculations \cite{freedman3}.\footnote{ This acquires a new importance in 
the light of the result in \cite{our3} that the divergent local terms of
the action contain information about the transformed generating
functional. To take the limit $\epsilon\rightarrow 0$ at this early stage 
would imply losing information.} This means that, instead of the equation 
above, we must consider

\beq
W_{\epsilon}[{\tilde f}_{\epsilon}]=\frac{\beta(\epsilon)}{2}\;
{\tilde f}_{\epsilon}^{2}\; ,
\label{18''}
\eeq
where we have replaced ${\tilde f}_{0}$ with ${\tilde
f}_{\epsilon}$. Note that there is still another important difference
between
Eqs.(\ref{18'}, \ref{18''}), namely, that we have also introduced a
dependence of the coupling $\beta$ on the distance to the boundary
$\epsilon$. We will show that this is needed for consistency.

From the considerations above, we write the perturbed functional as
\beq
I = \int d^{d}x
\; d^{d}y\;\sqrt{h}\int\frac{d^{d}k}{\left(
2\pi\right)^{d}}\;e^{-i\vec{k}\cdot\left(
\vec{x}-\vec{y}\right)}\left[\frac{1}{2}\;
\left(\frac{1}{F(k\epsilon)}+\beta (k\epsilon)\right)\;{\tilde
f}_{\epsilon}(\vec{x})\;{\tilde f}_{\epsilon}(\vec{y})\;+
f_{\epsilon}(\vec{x})
\;{\tilde f}_{\epsilon}(\vec{y})\right]\; ,
\label{19}
\eeq
which for $\beta =0$ reduces to Eq.(\ref{17}). The expression above is to
be contrasted with the one in \cite{muck}. There are two important
differences, namely, that $\beta$ depends on $\epsilon$, and that
$F(k\epsilon)$ contains all local and non-local terms, rather than only
the leading non-local term.\footnote{ The formulation in \cite{muck}
makes use, instead of $F(k\epsilon)$, of the conformal Green function
$G(k)=-\frac{\Gamma(1-\nu)}{\Gamma(1+\nu)}\;\left(\frac{k}{2}\right)^{2\nu}$, 
which involves a procedure in which the Legendre transformation is
performed after having selected the leading non-local term, rather
than following the opposite way as in \cite{our3}.}

Setting $\frac{\partial I}{\partial {\tilde f}_{\epsilon}}\; =0$ we
get

\beq
{\tilde
f}_{\epsilon}=-\frac{F(k\epsilon)}{1+\beta(k\epsilon)F(k\epsilon)}
\; f_{\epsilon}\; ,
\label{20}
\eeq
and introducing this into Eq.(\ref{19}) we find

\beq
I\left[f_{\epsilon}\right] = -\frac{1}{2}\int d^{d}x
\; d^{d}y\;\sqrt{h}\;f_{\epsilon}(\vec{x})
\;f_{\epsilon}(\vec{y})\int\frac{d^{d}k}{\left(
2\pi\right)^{d}}\;e^{-i\vec{k}\cdot\left(
\vec{x}-\vec{y}\right)}\;\frac{F(k\epsilon)}
{1+\beta(k\epsilon)F(k\epsilon)}\; .
\label{21}
\eeq
Comparison with Eq.(\ref{9}) shows that the double-trace perturbation
Eq.(\ref{18}) has
introduced the replacement

\beq
F(k\epsilon)\longrightarrow \frac{F(k\epsilon)}
{1+\beta(k\epsilon)F(k\epsilon)}\; .
\label{22}
\eeq
In the next section, we will show that this can be understood as
a modification of the
boundary conditions on the field, or equivalently, as the addition of a
surface term to the action Eq.(\ref{1}). We will also show how to extend
the formulation to the case of a scalar field non-minimally coupled to the
metric.

It has been stated in \cite{muck} that the formulation above applies to
both regular and irregular modes propagating in the bulk. However, the
formulation remains incomplete, because so far we have
only considered a perturbation in the conformal operator ${\cal O}$ (see
Eq.(\ref{18})). We still need to consider a perturbation in the conjugated
operator ${\tilde{\cal O}}$, and we will show that the
requirement for consistency imposes a precise constraint 
between both perturbations. 

Note that, in the transformed situation, we identify $f_{0}$ with
the conjugated operator ${\tilde{\cal O}}$. In order to do this, we
perform a Legendre transform in Eq.(\ref{9}) and get

\beq
{\tilde I}_{0} = -\int d^{d}x
\; d^{d}y\;\sqrt{h}\int\frac{d^{d}k}{\left(
2\pi\right)^{d}}\;e^{-i\vec{k}\cdot\left(
\vec{x}-\vec{y}\right)}\left[\frac{1}{2}\;F(k\epsilon)
\; f_{\epsilon}(\vec{x})
\; f_{\epsilon}(\vec{y})\;+
{\tilde f}_{\epsilon}(\vec{x})
\; f_{\epsilon}(\vec{y})\right]\; .
\label{24}
\eeq
We introduce the following double-trace perturbation in ${\tilde{\cal O}}$

\beq
{\tilde W}[{\tilde{\cal
O}}]=\frac{{\tilde\beta}}{2}\; {\tilde{\cal O}}^{2}\; ,
\label{23}
\eeq
and, identifying $f_{0}$ with     
the conjugated operator ${\tilde{\cal O}}$, we write the perturbed
functional as

\beq
{\tilde I} = -\int d^{d}x
\; d^{d}y\;\sqrt{h}\int\frac{d^{d}k}{\left(
2\pi\right)^{d}}\;e^{-i\vec{k}\cdot\left(
\vec{x}-\vec{y}\right)}\left[\frac{1}{2}\;
\left(F(k\epsilon)-{\tilde\beta} (k\epsilon)\right)
\; f_{\epsilon}(\vec{x})\; f_{\epsilon}(\vec{y})\;+
{\tilde f}_{\epsilon}(\vec{x})
\; f_{\epsilon}(\vec{y})\right]\; .
\label{25}
\eeq
Note that, as we have done with the coupling $\beta$, we have
introduced a dependence of ${\tilde\beta}$ on $\epsilon$. Setting
$\frac{\partial {\tilde I}}{\partial f_{\epsilon}}\; =0$ we get

\beq
f_{\epsilon}=-\frac{1}{F(k\epsilon)-{\tilde\beta}(k\epsilon)}\;
{\tilde f}_{\epsilon}\; ,
\label{26}
\eeq
and introducing this into Eq.(\ref{25}) we find

\beq
{\tilde I}\left[{\tilde f}_{\epsilon}\right] = \frac{1}{2}\int d^{d}x
\; d^{d}y\;\sqrt{h}\;{\tilde f}_{\epsilon}(\vec{x})
\;{\tilde f}_{\epsilon}(\vec{y})\int\frac{d^{d}k}{\left(
2\pi\right)^{d}}\;e^{-i\vec{k}\cdot\left(
\vec{x}-\vec{y}\right)}\;\frac{1}
{F(k\epsilon)-{\tilde\beta}(k\epsilon)}\; .
\label{27}
\eeq
There is still one last thing to be considered in order to have a
complete and consistent formulation, and it is to require the expression
above to be actually the Legendre transform of Eq.(\ref{21}). As
anticipated, this will impose a precise constraint between $\beta$ and
${\tilde\beta}$. By Legendre transforming Eq.(\ref{21}) we
find

\beq
{\tilde I}\left[{\tilde f}_{\epsilon}\right] = \frac{1}{2}\int d^{d}x
\; d^{d}y\;\sqrt{h}\;{\tilde f}_{\epsilon}(\vec{x})
\;{\tilde f}_{\epsilon}(\vec{y})\int\frac{d^{d}k}{\left( 
2\pi\right)^{d}}\;e^{-i\vec{k}\cdot\left(
\vec{x}-\vec{y}\right)}\;\frac{1+\beta(k\epsilon)F(k\epsilon)}
{F(k\epsilon)}\; .
\label{28}
\eeq
From Eqs.(\ref{27}, \ref{28}) we find the constraint

\beq
{\tilde\beta}(k\epsilon)=\frac{\beta(k\epsilon)F^{2}(k\epsilon)}
{1+\beta(k\epsilon)F(k\epsilon)}\; ,
\label{29}
\eeq
which is required for consistency. Note that both
functionals Eqs.(\ref{21}, \ref{28}) are needed in order to map to the
boundary all the information contained in the bulk. The corresponding
boundary CFT's can be obtained through the prescriptions Eqs.(\ref{13},
\ref{15}). We emphasize that the functionals Eqs.(\ref{21},
\ref{28}) contain all local and non-local terms, and that this is needed 
in order to map to the boundary the constraints that the quantization
imposes in the bulk \cite{our3}. This topic will be discussed in the
following section. 

So far, we have developed a generic formulation in which the double-trace
perturbations at the boundary of $AdS_{d+1}$ can be
consistently incorporated into an
extended AdS/CFT prescription. We still need to understand the precise way
in which such perturbations change the boundary conditions on the field 
by adding surface terms to the action Eq.(\ref{1}), and to extend the
formalism to the non-minimally coupled case. This is left for the
following section, where we will connect the formulation
developed so far to the generalized AdS/CFT prescription in \cite{our3},
which analyzes the role of boundary conditions in the AdS/CFT 
correspondence. We will also relate the previous formalism to the energy
of the theory on the bulk, and to the existence of constraints for
which the irregular modes are allowed to propagate. We will
find the explicit expressions of the couplings, and show that they exhibit
very interesting behaviors when irregular modes are also allowed to
propagate.

\section{Generalized Boundary Conditions}

In this section, we will show that the double-trace perturbations at the
boundary of $AdS_{d+1}$ can be understood as the introduction of
Dirichlet, Neumann and mixed boundary conditions on the scalar field. We
will consider both minimally and non-minimally
coupled cases. We begin by concentrating on the minimally coupled case.

\subsection{The Minimally Coupled Case}

In this case we set $\varrho =0$ in Eq.(\ref{11''}) thus getting $\nu
=\sqrt{\frac{d^{2}}{4}+m^{2}}$. It
is natural to begin by analyzing the simplest case in which the coupling
$\beta$ does not depend on the distance to the boundary. For this purpose,
we set

\beq
\beta(k\epsilon)=-2\lambda\; ,
\label{30}
\eeq
where $\lambda$ is a real coefficient. Then, from Eq.(\ref{29}) we also
find

\beq
{\tilde\beta}(k\epsilon)=-\frac{2\lambda F^{2}(k\epsilon)}{1-2\lambda 
F(k\epsilon)}\; .
\label{31}
\eeq
Note, then, that consistency of the formulation requires that at
least one of the couplings depends on the distance to the boundary, 
as anticipated in the previous section. We will later show that the
particular cases for which irregular modes are also allowed to propagate
in the bulk correspond to the situations in which $\beta$ and
${\tilde\beta}$ have different asymptotic behaviors. 

Under the identification Eq.(\ref{30}), the functionals Eqs.(\ref{21},
\ref{28}) read

\beq
I\left[f_{\epsilon}\right] = -\frac{1}{2}\int d^{d}x
\; d^{d}y\;\sqrt{h}\;f_{\epsilon}(\vec{x})
\;f_{\epsilon}(\vec{y})\int\frac{d^{d}k}{\left(
2\pi\right)^{d}}\;e^{-i\vec{k}\cdot\left(
\vec{x}-\vec{y}\right)}\;\frac{F(k\epsilon)}
{1-2\lambda F(k\epsilon)}\; ,
\label{32}
\eeq

\beq
{\tilde I}\left[{\tilde f}_{\epsilon}\right] = \frac{1}{2}\int d^{d}x
\; d^{d}y\;\sqrt{h}\;{\tilde f}_{\epsilon}(\vec{x})
\;{\tilde f}_{\epsilon}(\vec{y})\int\frac{d^{d}k}{\left(
2\pi\right)^{d}}\;e^{-i\vec{k}\cdot\left(
\vec{x}-\vec{y}\right)}\;\frac{1-2\lambda F(k\epsilon)}
{F(k\epsilon)}\; .
\label{33}
\eeq
The key result is that the functionals above are precisely the ones found
in \cite{our3} when considering a boundary condition which fixes at the
border the field

\beq
\phi + 2\lambda\;\partial_{n}\phi\; .
\label{34}
\eeq
Then, for the particular choice Eq.(\ref{30}), the double-trace 
perturbation acts by turning the usual Dirichlet boundary condition into a
mixed one.\footnote{ In \cite{our3}, this particular boundary condition
was
called as `Type II' mixed boundary condition.} This is also equivalent to
add to the action Eq.(\ref{1}) a boundary term of the form \cite{our3}

\beq
I=I_{0} +\;\lambda\int
d^{d}x\;\sqrt{h}\;\left(\partial_{n}\phi_{\epsilon}\right)^{2}\; .
\label{35}
\eeq
Note that, in the particular case $\lambda =0$, we recover the usual
Dirichlet boundary condition. One important point is that for general
$\lambda$ both functionals Eqs.(\ref{32}, \ref{33}) correspond to the
same boundary conformal dimension, namely $\Delta_{+}$. However, in the
particular situation in which both constraints \cite{our3}

\beq
\lambda = \frac{1}{2\Delta_{-}}\; ,
\label{36}
\eeq
and

\beq
\nu <1\; ,
\label{37}
\eeq
are satisfied, the divergent local terms of both functionals
Eqs.(\ref{32},
\ref{33}) cancel out. This fact encodes the information that the
Legendre transform interpolates between different conformal dimensions, 
namely $\Delta_{+}$ and $\Delta_{-}$. In this situation, the addition of
counterterms is not required. From the bulk point of view, in this case
both
regular and irregular modes can propagate, because the
canonical energy is conserved, positive and finite for both of them. When
Eq.(\ref{36}) is
satisfied, but Eq.(\ref{37}) is not, the conformal dimension $\Delta_{-}$
reaches
the unitarity bound $(d-2)/2$, and becomes independent of the mass. The
unitarity bound is also reached for $m=0$. For details, see \cite{our3}.

We will now show that there exists an interesting relation between the
values and asymptotic behaviors of $\beta$ and ${\tilde\beta}$ and the
phenomenon of the propagation of irregular modes in the bulk. Let us first
consider the situation in which Eq.(\ref{36}) is not satisfied. For
simplicity, we concentrate on the case of $\nu$ not
integer.\footnote{ The case of $\nu$ integer is analogous. The only
difference is that there arise additional logarithmic terms which must be
taken
into account (see, for example, \cite{our2}).} By expanding the Bessel
functions in Eq.(\ref{31}) in powers of $\epsilon$ we find

\beq
{\tilde\beta}(k\epsilon)=-\frac{2\lambda\Delta_{-}^{2}}{1-2\lambda
\Delta_{-}}+\cdots\; ,
\label{720}
\eeq
where the dots stand for higher order terms in $\epsilon$. Comparison
with Eq.(\ref{30}) shows that in this case, where only regular modes are
allowed to propagate in the bulk, $\beta$ and
${\tilde\beta}$ have the same asymptotic behavior, namely,
both of them differ from a constant only on terms which vanish as we
approach the boundary. We also note from Eq.(\ref{720}) that
${\tilde\beta}$ diverges as we get closer to the critical value
Eq.(\ref{36}) for which irregular modes are also allowed to
propagate. Note that this result is consistent with the statement in
\cite{witten7} that, as the coupling grows, the system approaches the
condition that is suitable for quantization to get a field of dimension
$\Delta_{-}$.

We finally consider the situation where Eq.(\ref{36}) is satisfied and 
irregular modes are allowed to propagate as well. We expect from the
results above that something special will happen in this particular
situation. We will show that, in fact, in this case $\beta$ and
${\tilde\beta}$ have different asymptotic behaviors, and the last one
diverges as we approach the boundary. We have to consider separately
the cases where Eq.(\ref{37}) is, or not, also satisfied. Expanding in
Eq.(\ref{31}) we find for $\nu <1$  

\beq
{\tilde\beta}(k\epsilon)=-2^{2\nu
-1}\Delta_{-}^{2}\;\frac{\Gamma(\nu)}{\Gamma(1-\nu)}\;
(k\epsilon)^{-2\nu}+\cdots\; ,
\label{721}
\eeq
whereas for $\nu >1$ we get

\beq
{\tilde\beta}(k\epsilon)=-2(\nu
-1)\Delta_{-}^{2}(k\epsilon)^{-2}+\cdots\; .
\label{722}
\eeq
In the equations above, the dots stand for higher order terms in
$\epsilon$. As anticipated we find that, in the particular
situation where irregular modes are also allowed to propagate in the bulk,
the couplings $\beta$ and ${\tilde\beta}$ have different asymptotic
behaviors, and the last one, which corresponds to the operator of
conformal dimension $\Delta_{+}$ \cite{our3}, diverges as we approach the
boundary.

So far, we have considered the situation where $\beta$ does
not
depend on the distance to the boundary. As the next step, it is natural to
analyze the case where $\beta$ depends on $\epsilon$,
and ${\tilde\beta}$ does not. In order to do this, we consider, instead of
Eq.(\ref{30}),
the following starting point

\beq
{\tilde\beta}(k\epsilon)=-2\lambda\; ,
\label{723}
\eeq
where, as before, $\lambda$ is a real coefficient. From Eq.(\ref{29}) we
get

\beq
\beta(k\epsilon)=-\frac{2\lambda}
{F(k\epsilon)}\;\frac{1}{F(k\epsilon)+2\lambda}\; .
\label{724}
\eeq

From the equation above, we write the functionals Eqs.(\ref{21}, 
\ref{28}) as

\ba
I\left[f_{\epsilon}\right] &=& -\frac{1}{2}\int d^{d}x
\; d^{d}y\;\sqrt{h}\;f_{\epsilon}(\vec{x})
\;f_{\epsilon}(\vec{y})\int\frac{d^{d}k}{\left(
2\pi\right)^{d}}\;e^{-i\vec{k}\cdot\left(
\vec{x}-\vec{y}\right)}\;\left[ F(k\epsilon)+2\lambda\right]\; ,
\label{725}\\ \nonumber\\ \nonumber\\
{\tilde I}\left[{\tilde f}_{\epsilon}\right] &=& \frac{1}{2}\int d^{d}x 
\; d^{d}y\;\sqrt{h}\;{\tilde f}_{\epsilon}(\vec{x})
\;{\tilde f}_{\epsilon}(\vec{y})\int\frac{d^{d}k}{\left(   
2\pi\right)^{d}}\;e^{-i\vec{k}\cdot\left(
\vec{x}-\vec{y}\right)}\;\frac{1}
{F(k\epsilon)+2\lambda}\; .
\label{726}
\ea
The functionals above have not been considered in \cite{our3}. However, 
by following an analogous procedure, it can be verified that
they correspond to consider a Dirichlet boundary condition and
add to the usual action Eq.(\ref{1}) a boundary term of the form

\beq
I=I_{0} -\;\lambda\int
d^{d}x\;\sqrt{h}\; \phi^{2}_{\epsilon}\; .
\label{727}
\eeq
The fact that the action above is stationary under a Dirichlet
boundary condition can be verified by noting that, under a
transformation $\phi\rightarrow\phi +\delta\phi$, its variation is given 
by the following surface term

\beq
\delta I =\int
d^{d}x\;\sqrt{h}\;(\partial_{n}\phi_{\epsilon}-2\lambda\phi_{\epsilon})
\;\delta
\phi_{\epsilon}\; ,
\label{728}
\eeq
which vanishes for $\delta\phi_{\epsilon}=0$. Since we are still
considering a Dirichlet boundary condition, it may seem that in this
situation the double-trace perturbation causes no effect. However, this is
not the case, because the importance of the surface term in
Eq.(\ref{727}) is that it
allows for irregular modes to propagate in the bulk for particular values
of $\lambda$. This can be verified by making use of a formalism
which is analogous to the one developed in \cite{our3} for other possible
boundary conditions. In particular, it can be seen that for

\beq
\lambda =-\frac{\Delta_{-}}{2}\; ,
\label{729}
\eeq
the canonical energy is conserved, positive and finite for both regular
and irregular modes propagating in the bulk, provided that the constraint
Eq.(\ref{37}) is satisfied as well.\footnote{Since this particular case 
has not been analyzed in \cite{our3}, we summarize here the
main results. Making use of global coordinates
$(\tau,\rho,\vec{\Omega})$ defined as in \cite{our3} (recall here that
$\tau$ is
the time coordinate, $\rho$ is the radial coordinate and $\vec{\Omega}$
are the angular coordinates) we find from the action Eq.(\ref{727}) that
the canonical energy is of the form $E=-\int
d^{d}x\;\sqrt{g}\;\left[\Theta^{\tau}_{\;\tau}+\lambda
\nabla_{\mu}(n^{\mu}\phi^{2})\right]$, 
where the integration is carried out over the spatial coordinates, and
$\Theta_{\mu\nu} =
\partial_{\mu}\phi\;\partial_{\nu}\phi\;-\;\frac{1}{2}
\; g_{\mu\nu}\left[
g^{\alpha\beta}\partial_{\alpha}\phi\;\partial_{\beta}\phi\;
+m^{2}\phi^{2}\right]$. It
can be verified that, for regular modes propagating in the bulk, the
canonical energy is conserved, positive and finite for any values of
$\lambda$ and $\nu$, whereas for irregular modes propagating in the
bulk the canonical energy is conserved, positive and finite only when the 
constraints Eqs.(\ref{37}, \ref{729}) are satisfied.} We emphasize that,
when both constraints Eqs.(\ref{37}, \ref{729}) are satisfied, the
Legendre transform interpolates between different conformal dimensions
$\Delta_{+}$ and $\Delta_{-}$, due to the cancellation of the divergent
local terms of the functionals Eqs.(\ref{725}, \ref{726}). This
statement can be verified by following a procedure analogous to the one in
\cite{our3}. When
Eq.(\ref{729}) is satisfied, but Eq.(\ref{37}) is not, the conformal
dimension $\Delta_{-}$ reaches the unitarity bound $(d-2)/2$.

Now we analyze the existing relation between the values and
asymptotic
behaviors of $\beta$ and ${\tilde\beta}$ and the phenomenon of the
propagation of irregular modes. We have seen that in the previously
analyzed mixed boundary condition the coupling ${\tilde\beta}$ diverges
when the
critical value Eq.(\ref{36}) is reached. Since in the current Dirichlet 
situation the conformal dimension $\Delta_{-}$ corresponds to the conformal
operator ${\tilde{\cal O}}$, rather than to 
${\cal O}$,\footnote{This 
result can be found by computing the boundary CFT's through the 
prescription Eqs.(\ref{13}, \ref{15})} then in
this situation we expect that it will
be $\beta$,
rather than ${\tilde\beta}$, the coupling which diverges as we
approach the critical value Eq.(\ref{729}). We now verify this
statement. Again, we concentrate on the case of $\nu$ not
integer. Let us first consider the situation when Eq.(\ref{729}) is not
satisfied. Expanding in $\epsilon$ in Eq.(\ref{724}) we find

\beq
\beta(k\epsilon)=-\frac{2\lambda}{\Delta_{-}(\Delta_{-}+2\lambda)}+\cdots\; ,
\label{730}
\eeq
where the dots stand for higher order terms in $\epsilon$. We note from
Eq.(\ref{723}) that, as in the previous mixed case, $\beta$ and
${\tilde\beta}$ have the same asymptotic behavior when only regular modes
propagate in the bulk. Note also that, as
expected, $\beta$ diverges as we approach the critical value
Eq.(\ref{729}) for which irregular modes are allowed to propagate as well.

Now we analyze the case when the constraint Eq.(\ref{729}) is
satisfied. The calculations are analogous to the ones performed in the
previous mixed case, and the results are as follows. For $\nu <1$ we get

\beq
\beta(k\epsilon)=2^{2\nu}\frac{\lambda}{\Delta_{-}}
\;\frac{\Gamma(\nu)}{\Gamma(1-\nu)}\;
(k\epsilon)^{-2\nu}+\cdots\; ,
\label{731}
\eeq 
whereas for $\nu >1$ we find

\beq
\beta(k\epsilon)=4(\nu
-1)\;\frac{\lambda}{\Delta_{-}}\;(k\epsilon)^{-2}+\cdots\; .
\label{732}
\eeq
Then, as in the previous mixed case, in the situation when the irregular
modes are also allowed to propagate the couplings have different
asymptotic behaviors, as noticed by comparing Eqs.(\ref{731},
\ref{732}) to Eq.(\ref{723}). We also note that in this case the coupling
$\beta$, which corresponds to the conformal operator of dimension
$\Delta_{+}$, diverges as we approach the boundary.

So far, we have considered the situations in which one of the couplings
does
not depend on $\epsilon$. Since the consistency relation
Eq.(\ref{29}) makes it impossible to consider a case where both
couplings are simultaneously independent of the distance to the
boundary, there is only
one case left to be considered, namely, the situation in which both
couplings depend on $\epsilon$. In order to analyze this case, we set

\beq
\beta(k\epsilon)=-\frac{1+F(k\epsilon)\left[F(k\epsilon)-2
\lambda\right]}{F(k\epsilon)}\; ,
\label{733}
\eeq
and from Eq.(\ref{29}) we also find

\beq
{\tilde\beta}(k\epsilon)=\frac{1+F(k\epsilon)\left[F(k\epsilon)-2\lambda
\right]}{F(k\epsilon)-2\lambda}\; .
\label{734}
\eeq
The reason why we choose the particular expressions above will be
clarified later. Introducing Eq.(\ref{733}) into Eqs.(\ref{21},
\ref{28}) we get

\ba
I\left[f_{\epsilon}\right] &=& \frac{1}{2}\int d^{d}x
\; d^{d}y\;\sqrt{h}\;f_{\epsilon}(\vec{x})
\;f_{\epsilon}(\vec{y})\int\frac{d^{d}k}{\left(
2\pi\right)^{d}}\;e^{-i\vec{k}\cdot\left(
\vec{x}-\vec{y}\right)}\;\frac{1}{F(k\epsilon)-2\lambda}\; ,
\label{735}
\\ \nonumber\\ \nonumber\\
{\tilde I}\left[{\tilde f}_{\epsilon}\right] &=& -\frac{1}{2}\int d^{d}x
\; d^{d}y\;\sqrt{h}\;{\tilde f}_{\epsilon}(\vec{x})
\;{\tilde f}_{\epsilon}(\vec{y})\int\frac{d^{d}k}{\left(
2\pi\right)^{d}}\;e^{-i\vec{k}\cdot\left(
\vec{x}-\vec{y}\right)}\;\left[F(k\epsilon)-2\lambda\right]\; .
\label{736}
\ea
The key result is that, as it can be verified, the functionals above 
correspond to add to the action Eq.(\ref{1}) the following surface terms

\beq
I=I_{0} -\;\lambda\int
d^{d}x\;\sqrt{h}\; \phi^{2}_{\epsilon}- \int
d^{d}x\;\sqrt{h}\;\phi_{\epsilon}\;\partial_{n}\phi_{\epsilon}\; ,
\label{737}   
\eeq
which give rise to an action which is stationary under a boundary
condition which fixes at the border the field \footnote{Note that, under
a transformation $\phi\rightarrow\phi +\delta\phi$, the variation of the
action Eq.(\ref{737}) is given by the surface term $\delta I=-\int
d^{d}x\;\sqrt{h}\;\phi_{\epsilon}\;\delta (\partial_{n}\phi_{\epsilon} +
2\lambda\phi_{\epsilon})$ which
vanishes under a boundary condition which fixes at the border the field 
Eq.(\ref{738}).}

\beq
\partial_{n}\phi + 2\lambda\phi\; .
\label{738}
\eeq
This means that the double-trace perturbation Eq.(\ref{733}) turns the
usual Dirichlet boundary condition into a mixed one which is different
from the previously analyzed mixed boundary condition (see
Eq.(\ref{34})). This clarifies the reason why we chose the specific
coupling
Eq.(\ref{733}). Different choices may not make sense in terms of a
problem of boundary conditions on the scalar field. Note that, in
the particular case $\lambda =0$, this new mixed boundary condition
reduces to a Neumann one. Even when this boundary condition has not been
analyzed in \cite{our3}, the procedure is analogous, and it can be
verified that in this situation the irregular modes are allowed to
propagate in the bulk as well as the regular ones only when the following
constraint \footnote{ Since this particular case has not been 
analyzed in \cite{our3}, we give here the main results. The
canonical energy corresponding to the action Eq.(\ref{737}) is given by
$E=-\int d^{d}x\;\sqrt{g}\;\left[\Theta^{\tau}_{\;\tau}+\lambda
\nabla_{\mu}(n^{\mu}\phi^{2})+\frac{1}{2}\;\left(\nabla^{2}\phi^{2}-
\partial^{\tau}\partial_{\tau}\phi^{2}\right)\right]$,
where we consider global
coordinates $(\tau,\rho,\vec{\Omega})$ defined as in \cite{our3}, the
integration is carried out over the spatial coordinates, and
$\Theta_{\mu\nu} =
\partial_{\mu}\phi\;\partial_{\nu}\phi\;-\;\frac{1}{2}
\; g_{\mu\nu}\left[
g^{\alpha\beta}\partial_{\alpha}\phi\;\partial_{\beta}\phi\;
+m^{2}\phi^{2}\right]$. It can be shown that the canonical energy is
conserved, positive and finite for irregular modes propagating in the
bulk only when both constraints Eqs.(\ref{37}, \ref{739}) are satisfied.}

\beq
\lambda =\frac{\Delta_{-}}{2}\; ,
\label{739}
\eeq
is satisfied together with Eq.(\ref{37}). It can be shown that in this
situation the divergent local terms of the functionals
Eqs.(\ref{735}, \ref{736}) cancel out and,
correspondingly, the Legendre transform interpolates between different
conformal dimensions, namely $\Delta_{+}$ and $\Delta_{-}$. When
Eq.(\ref{739}) is satisfied, but Eq.(\ref{37}) is not, the conformal
dimension $\Delta_{-}$ reaches the unitarity bound $(d-2)/2$.

Regarding the relation between the values and asymptotic
behaviors of $\beta$ and ${\tilde\beta}$ and the phenomenon of the
propagation of irregular modes, we expect to find results which are
analogous to the ones found in the previous cases. In particular, we
expect that in this situation the coupling ${\tilde\beta}$, which is the
one that
corresponds to the conformal operator of dimension $\Delta_{+}$, 
\footnote{ This 
result can be found by computing the boundary CFT's through the 
prescription Eqs.(\ref{13}, \ref{15})} should
diverge as we approach the critical value Eq.(\ref{739}). Now we verify
this statement for the case of $\nu$ not integer. Since calculations are
analogous to the ones performed in the previous cases, we just present the
main results. Let us first consider that Eq.(\ref{739}) is not
satisfied. Then, expanding in Eqs.(\ref{733}, \ref{734}) we find

\ba
\beta(k\epsilon)&=&-\frac{1}{\Delta_{-}}\;\left[1+
\Delta_{-}(\Delta_{-}-2\lambda)\right]+\cdots\; ,
\label{740}\\
{\tilde\beta}(k\epsilon)&=&\frac{1}{\Delta_{-}-2\lambda}\;\left[1+
\Delta_{-}(\Delta_{-}-2\lambda)\right]+\cdots\; ,
\label{741}
\ea
where, as in the previous cases, the dots stand for higher order terms in
$\epsilon$. As anticipated, we note that $\beta$ and ${\tilde\beta}$ have
the same
asymptotic behavior, and that ${\tilde\beta}$ diverges as we approach the
critical value Eq.(\ref{739}).

Finally, we consider the situation in which the constraint
Eq.(\ref{739}) is satisfied. For $\nu <1$ we find

\ba
\beta(k\epsilon)&=&-\frac{1}{\Delta_{-}}\; + \cdots\; ,
\label{742}\\
{\tilde\beta}(k\epsilon)&=&-2^{2\nu
-1}\;\frac{\Gamma(\nu)}{\Gamma(1-\nu)}\;
(k\epsilon)^{-2\nu}+\cdots\; ,
\label{743}
\ea
whereas for $\nu >1$ we get

\ba
\beta(k\epsilon)&=&-\frac{1}{\Delta_{-}}\; + \cdots\; ,
\label{744}\\
{\tilde\beta}(k\epsilon)&=&-2(\nu
-1)(k\epsilon)^{-2}+\cdots\; .
\label{745}
\ea
Then, as expected, the couplings have different asymptotic behaviors, and
${\tilde\beta}$ diverges as we approach the boundary.

\subsection{The Non-Minimally Coupled Case}

In the previous sub-section, we have shown, for the particular minimally
coupled case, 
that the perturbation of the conformal field theory by a double-trace
operator can be understood as the introduction of a generalized boundary
condition on the scalar field. Now we consider the more general situation
of a non-minimally coupled scalar field. In this case, we have a
non-vanishing coupling coefficient $\varrho$ in Eq.(\ref{11''}). But, as
pointed out in \cite{our3}, the effect of the non-minimal coupling does
not limit itself to a redefinition of the effective mass of the
theory. In this subsection we will show, in particular, that the
introduction of a double-trace perturbation at the boundary generates the
natural extension of the Gibbons-Hawking surface term \cite{gibbons},
which is added to the Einstein-Hilbert action in order to
have a well-defined variational principle under variations of the
metric. As in the minimally coupled case, we will consider different
boundary
conditions on the scalar field and perform a detailed analysis of
each one of them.

We begin by considering the simplest case where the coupling
${\tilde\beta}$
does not depend on the distance to the boundary

\beq
{\tilde\beta}(k\epsilon)=2\varrho d\; ,
\label{49}
\eeq
where $\varrho$ has been introduced in Eq.(\ref{11''}). Recall here that,
throughout this section, we will consider
$\nu=\sqrt{\frac{d^2}{4}\;+\;m^{2}+\varrho R}$. From Eq.(\ref{29}) we also
find

\beq
\beta(k\epsilon)=\frac{2\varrho
d}{F(k\epsilon)}\;\frac{1}{F(k\epsilon)-2\varrho d}\; .
\label{48}
\eeq

Introducing the equation above into Eqs.(\ref{21}, \ref{28}) we get

\ba
I\left[f_{\epsilon}\right] &=& -\frac{1}{2}\int d^{d}x
\; d^{d}y\;\sqrt{h}\;f_{\epsilon}(\vec{x})
\;f_{\epsilon}(\vec{y})\int\frac{d^{d}k}{\left(
2\pi\right)^{d}}\;e^{-i\vec{k}\cdot\left(
\vec{x}-\vec{y}\right)}\;\left[ F(k\epsilon)-2\varrho d\right]\; ,
\label{50}\\ \nonumber\\ \nonumber\\
{\tilde I}\left[{\tilde f}_{\epsilon}\right] &=& \frac{1}{2}\int d^{d}x  
\; d^{d}y\;\sqrt{h}\;{\tilde f}_{\epsilon}(\vec{x})
\;{\tilde f}_{\epsilon}(\vec{y})\int\frac{d^{d}k}{\left(
2\pi\right)^{d}}\;e^{-i\vec{k}\cdot\left(
\vec{x}-\vec{y}\right)}\;\frac{1}
{F(k\epsilon)-2\varrho d}\; .
\label{51}
\ea
The functionals above are precisely the same as the ones found in
\cite{our3} when considering, in the non-minimally coupled case, a
Dirichlet boundary condition which fixes the field $\phi$ at the
border. We note from \cite{our3} that the functionals above correspond to
add to the usual action Eq.(\ref{1}) a boundary term of the form 

\beq
I=I_{0} -\;\varrho\int
d^{d}x\;\sqrt{h}\; {\cal K}_{\epsilon}\;\phi^{2}_{\epsilon}\; ,
\label{52}
\eeq
where ${\cal K}$ is the trace of the extrinsic curvature at the
boundary. The above
surface term is just the natural extension of the usual Gibbons-Hawking
term \cite{gibbons}. The
importance of this surface term is that, even when we are considering
a Dirichlet boundary condition, it lets irregular modes to propagate in
the bulk for particular values of $\varrho$ \cite{our3}. We have just
shown that such surface term can be generated by perturbing the conformal
field theory with a double-trace operator. Note that in the particular
case $\varrho =0$, corresponding to the minimally coupled case, we recover
the usual Dirichlet boundary condition. We also point out that for general
$\varrho$ both functionals Eqs.(\ref{50}, \ref{51}) give rise to the same
conformal dimension $\Delta_{+}$, with one particular exception,
namely, the situation in which both constraints
Eq.(\ref{37}) and \cite{our3}

\beq
\varrho=\frac{d-1}{8d}\;
\left[1\pm\sqrt{1+\left(\frac{4m}{d-1}\right)^{2}}\right]\; ,
\label{53}
\eeq
are satisfied.\footnote{ It is interesting to note that, in the
particular case $m=0$, one of the solutions in Eq.(\ref{53}) vanishes,
whereas the another one reduces to the conformal value for which the
`improved' stress-energy tensor of the scalar field becomes
traceless. This
means that, in particular, Weyl-invariant theories in the bulk allow
for irregular modes to propagate as well, and give rise to functionals
in which the divergent local terms cancel out.} In this case, the
divergent
local terms of both functionals
Eqs.(\ref{50}, \ref{51}) cancel out, making the addition of
counterterms to be unnecessary, and the Legendre transform
interpolates between different conformal
dimensions, namely $\Delta_{+}$ and $\Delta_{-}$. From the bulk point of
view, both regular and irregular modes can propagate, because the
canonical energy is conserved, positive and finite for both of them. When
Eq.(\ref{53}) is satisfied but
Eq.(\ref{37}) is not, the conformal dimension $\Delta_{-}$
reaches the unitarity bound $(d-2)/2$. For details, see \cite{our3}.

Now we concentrate on the relation between the
values and asymptotic behaviors of $\beta$ and ${\tilde\beta}$ and the
phenomenon of the propagation of irregular modes in the bulk. Note that, 
in this case, the conformal dimension $\Delta_{-}$ corresponds to the
conformal operator ${\tilde{\cal O}}$ \cite{our3}, and we expect $\beta$ 
to diverge as we
approach any of the critical values Eq.(\ref{53}). We begin by considering
the situation in which Eq.(\ref{53}) is not satisfied. Expanding for $\nu$
not integer in Eq.(\ref{48}) we find

\beq
\beta(k\epsilon)=\frac{2\varrho 
d}{\Delta_{-}(\Delta_{-}-2\varrho d)}+\cdots\; ,
\label{2300}
\eeq
where the dots stand for higher order terms in $\epsilon$. From
Eqs.(\ref{49}, \ref{2300}) we find that $\beta$ and
${\tilde\beta}$ have the same asymptotic behavior, as expected. Also as
expected, we note from the equation above that $\beta$ diverges as we
approach any of the critical values Eq.(\ref{53}).\footnote{ Recall here
that $\Delta_{-}$ depends on $\varrho$.}

We still need to consider the case when Eq.(\ref{53}) is satisfied. This
is done as in the previous cases. For $\nu <1$ we get

\beq
\beta(k\epsilon)=-2^{2\nu}\frac{\varrho d}{\Delta_{-}}
\;\frac{\Gamma(\nu)}{\Gamma(1-\nu)}\;
(k\epsilon)^{-2\nu}+\cdots\; ,
\label{2301}
\eeq
whereas for $\nu >1$ we find

\beq
\beta(k\epsilon)=-4(\nu
-1)\;\frac{\varrho d}{\Delta_{-}}\;(k\epsilon)^{-2}+\cdots\; .
\label{2302}
\eeq
As expected, if irregular modes are also allowed to propagate, then the
couplings have different asymptotic behaviors, as noticed from
Eqs.(\ref{49}, \ref{2301}, \ref{2302}). In this situation, $\beta$
diverges
as we approach the boundary.

Finally, we consider a
situation in which both couplings $\beta$ and
${\tilde\beta}$ depend on the distance to the boundary. This is done by
setting

\beq
\beta(k\epsilon)=-\frac{1+F(k\epsilon)\left[F(k\epsilon)+2\varrho 
d\right]}{F(k\epsilon)}\; .
\label{55}
\eeq
From Eq.(\ref{29}) we also get

\beq
{\tilde\beta}(k\epsilon)=\frac{1+F(k\epsilon)\left[F(k\epsilon)+2\varrho
d\right]}{F(k\epsilon)+2\varrho d}\; .
\label{56}
\eeq
Introducing Eq.(\ref{55}) into Eqs.(\ref{21}, \ref{28}) we find

\ba
I\left[f_{\epsilon}\right] &=& \frac{1}{2}\int d^{d}x
\; d^{d}y\;\sqrt{h}\;f_{\epsilon}(\vec{x})
\;f_{\epsilon}(\vec{y})\int\frac{d^{d}k}{\left(
2\pi\right)^{d}}\;e^{-i\vec{k}\cdot\left(
\vec{x}-\vec{y}\right)}\;\frac{1}{F(k\epsilon)+2\varrho
d}\; ,
\label{57}
\\ \nonumber\\ \nonumber\\
{\tilde I}\left[{\tilde f}_{\epsilon}\right] &=& -\frac{1}{2}\int d^{d}x
\; d^{d}y\;\sqrt{h}\;{\tilde f}_{\epsilon}(\vec{x})
\;{\tilde f}_{\epsilon}(\vec{y})\int\frac{d^{d}k}{\left(
2\pi\right)^{d}}\;e^{-i\vec{k}\cdot\left(
\vec{x}-\vec{y}\right)}\;\left[F(k\epsilon)+2\varrho d\right]\; .
\label{58}
\ea
The key result is that the functionals above are precisely the same as the
ones considered in \cite{our3} when analyzing a boundary condition which
fixes at the boundary the field

\beq
\partial_{n}\phi + 2\varrho {\cal K}\phi\; .
\label{59}
\eeq
This means that, for the particular choice Eq.(\ref{55}), the double-trace
perturbation acts by turning the Dirichlet boundary condition into a mixed
one, which is different from the ones considered in the minimally coupled
case (see Eqs.(\ref{34}, \ref{738})).\footnote{ In \cite{our3}, this
particular
boundary condition was called as `Type I' mixed boundary condition.} It
is also interesting to note from Eq.(\ref{59}) that, in the particular
minimally coupled
case ($\varrho =0$), this mixed boundary condition reduces to a Neumann
one. We also point out that setting this mixed boundary condition is
equivalent to add to the usual action Eq.(\ref{1}) the following boundary
terms \cite{our3}

\beq  
I=I_{0} -\;\varrho\int
d^{d}x\;\sqrt{h}\; {\cal K}_{\epsilon}\;\phi^{2}_{\epsilon}- \int
d^{d}x\;\sqrt{h}\;\phi_{\epsilon}\;\partial_{n}\phi_{\epsilon}\; ,
\label{60}
\eeq
where the first surface term is the same that arises in the Dirichlet
situation Eq.(\ref{52}), whereas the second surface term is new. We
emphasize that the new surface term does not spoil the property of having
a well-defined variational principle under variations of the metric. We
also point out that for general
$\varrho$ both functionals Eqs.(\ref{57}, \ref{58}) give rise to the
same conformal dimension $\Delta_{+}$, with one particular exception,   
namely the situation in which both constraints Eq.(\ref{37}) and
\cite{our3}

\beq
\varrho=-\frac{3d+1}{8d}\;
\left[1\mp\sqrt{1+\left(\frac{4m}{3d+1}\right)^{2}}\right]\; ,
\label{61}
\eeq
are satisfied. In analogy to the former cases, in this situation the
divergent local terms of the functionals Eqs.(\ref{57}, \ref{58}) cancel
out, and this fact encodes the information
that the Legendre transform interpolates between different conformal   
dimensions $\Delta_{+}$ and $\Delta_{-}$. In this case there is no
need to add any counterterms. From the bulk point of
view, both regular and irregular modes are allowed to propagate, because
the
canonical energy is conserved, positive and finite for both of them. In
addition, when
Eq.(\ref{61}) is satisfied but Eq.(\ref{37}) is not, the conformal
dimension $\Delta_{-}$
reaches the unitarity bound $(d-2)/2$ and becomes independent of the
effective mass. For details, see \cite{our3}.

We still need to analyze the relation between the values and asymptotic
behaviors of $\beta$ and ${\tilde\beta}$ and the phenomenon of the
propagation of irregular modes. In this particular case, we expect 
${\tilde\beta}$ to diverge as we approach any of the critical values
Eq.(\ref{61}). This is due to the fact that ${\tilde\beta}$ corresponds to
the conformal operator of dimension $\Delta_{+}$ \cite{our3}. Let us first
consider the situation where Eq.(\ref{61}) is not satisfied. Expanding in
Eqs.(\ref{55}, \ref{56}) we get

\ba
\beta(k\epsilon)&=&-\frac{1}{\Delta_{-}}\;\left[1+
\Delta_{-}(\Delta_{-}+2\varrho d)\right]+\cdots\; ,
\label{2700}\\
{\tilde\beta}(k\epsilon)&=&\frac{1}{\Delta_{-}+2\varrho d}\;\left[1+
\Delta_{-}(\Delta_{-}+2\varrho d)\right]+\cdots\; .
\label{2701}
\ea
As expected, $\beta$ and ${\tilde\beta}$ have the same
asymptotic behavior, and ${\tilde\beta}$ diverges as we approach
any of the critical values Eq.(\ref{61}).\footnote{ Recall here   
that $\Delta_{-}$ depends on $\varrho$.}

Finally, we consider the situation in which Eq.(\ref{61}) is
satisfied. For $\nu <1$ we get

\ba
\beta(k\epsilon)&=&-\frac{1}{\Delta_{-}}\; + \cdots\; ,
\label{2705}\\
{\tilde\beta}(k\epsilon)&=&-2^{2\nu
-1}\;\frac{\Gamma(\nu)}{\Gamma(1-\nu)}\;
(k\epsilon)^{-2\nu}+\cdots\; ,
\label{2702}
\ea
whereas for $\nu >1$ we have

\ba
\beta(k\epsilon)&=&-\frac{1}{\Delta_{-}}\; + \cdots\; ,
\label{2706}\\
{\tilde\beta}(k\epsilon)&=&-2(\nu
-1)(k\epsilon)^{-2}+\cdots\; .
\label{2703}
\ea
As expected, $\beta$ and ${\tilde\beta}$ have different asymptotic
behaviors, and
${\tilde\beta}$ diverges as we approach the boundary.

\section{Conclusions}

In this work, we have introduced a generalized AdS/CFT prescription, 
which is suggested by results in \cite{witten7}\cite{berkooz2}\cite{our3}, 
and can consistently incorporate double-trace perturbations at the 
boundary CFT. This leads to a formalism in which we can perform explicit 
calculations. We have analyzed both minimally and non-minimally coupled 
cases, and obtained new results for both of them. 

We have shown that consistency of the formalism imposes a precise relation
between the couplings $\beta$ and ${\tilde\beta}$ (see
Eq.(\ref{29})), and requires that at least one of them depends on the
distance to the boundary. We have considered many possible couplings, and
in all situations we have shown that, to introduce a double-trace
perturbation at the border, is equivalent to add a surface term to the
usual action Eq.(\ref{1}). This gives rise to different kinds of
Dirichlet, Neumann and mixed boundary conditions on the scalar field. In 
all cases, we have performed the explicit calculation of $\beta$ and
${\tilde\beta}$ and of the corresponding surface terms. We have found that
there exist particular values of $\beta$ and ${\tilde\beta}$ which allow 
for irregular modes to propagate as well. We have computed the explicit 
expressions of such special couplings. We have shown that,
as we get closer to the situation in
which irregular modes are allowed to propagate, the coupling corresponding
to the operator of conformal dimension $\Delta_{+}$ diverges. This result
is
consistent with the statement in \cite{witten7} that, as the coupling
grows, the system approaches the condition that is suitable for
quantization to get a field of dimension $\Delta_{-}$. In
addition, we have shown that, when the constraints for which irregular
modes propagate are satisfied, $\beta$ and ${\tilde\beta}$
have different asymptotic behaviors, and the coupling corresponding to the
operator of conformal dimension $\Delta_{+}$ diverges as we approach the
boundary. 

We have also shown that there exist particular values of $\beta$ and
${\tilde\beta}$ which require the introduction of new boundary conditions
which have not been considered in \cite{our3}. We have included these new
boundary conditions in our analysis.

In the particular non-minimally coupled case, we have also shown that the
introduction of a double-trace perturbation at the boundary generates the
natural extension of the Gibbons-Hawking surface term.

In general, we have shown that the introduction of double-trace
perturbations can be understood in terms of the
formulation in \cite{our3}. In particular, this enables us to relate the
double-trace perturbations to the quantization which makes use of the
canonical energy instead of the metrical one. Such quantization appears 
to be the natural one to be considered in the AdS/CFT context.

Throughout this article, we have only considered the simplest 
non-trivial case of multi-trace perturbations, namely, that of 
double-trace ones. It would be interesting to extend this
formalism to the case of higher power trace perturbations. It 
should be noted that in such case non-linear boundary conditions would 
arise which would require to consider a perturbative approach. More 
complex calculations would be involved to first extend the formulation 
in \cite{our3} and then to employ it to describe the more complicated 
multi-trace perturbations. We also would like to be able to reproduce the 
constraints for which irregular modes propagate in the bulk and the 
couplings diverge by performing calculations exclusively on the boundary 
CFT side. It also would be interesting to understand the meaning of the 
generalized boundary conditions in the string theory context. This 
requires a more detailed study.

\section{Acknowledgements}

I would like to thank J. Maldacena for helpful discussions. Special
thanks to V. O. Rivelles for encouragement and a critical reading of
the manuscript. This work was supported by FAPESP grant 01/05770-1.


\begin{thebibliography}{99}
\bibitem{berkooz1} O. Aharony, M. Berkooz, and E. Silverstein,
``Multiple-Trace Operators and Non-Local String Theories'', JHEP {\bf
0108} (2001) 006, hep-th/0105309.
\bibitem{berkooz3} O. Aharony, M. Berkooz and E. Silverstein, ``Non-local
string theories on $AdS_{3}\times S^{3}$ and stable non-supersymmetric
backgrounds'', hep-th/0112178.
\bibitem{maldacena} J. M. Maldacena, ``The Large N Limit of Superconformal
Field Theories and Supergravity'', Adv. Theor. Math. Phys. {\bf 2}
(1998) 231, hep-th/9711200.
\bibitem{mald} O. Aharony, S. S. Gubser, J. Maldacena, H. Ooguri and
Y. Oz,
``Large N Field Theories, String Theory and Gravity'', Phys. Rept. {\bf
323} (2000) 183, hep-th/9905111.
\bibitem{freedman9} E. D'Hoker, S. D. Mathur, A. Matusis and L. Rastelli,
``The Operator Product Expansion of N=4 SYM and the 4-point Functions of
Supergravity'', Nucl. Phys. {\bf B589} (2000) 38, hep-th/9911222. 
\bibitem{witten7} E. Witten, ``Multi-Trace Operators, Boundary Conditions,
And AdS/CFT Correspondence'', hep-th/0112258.
\bibitem{berkooz2} M. Berkooz, A. Sever and A. Shomer,
``"Double-trace" Deformations, Boundary Conditions and Spacetime
Singularities'', hep-th/0112264.
\bibitem{witten} E. Witten, ``Anti De Sitter Space and Holography'',  
Adv. Theor. Math. Phys. {\bf 2} (1998) 253, hep-th/9802150.
\bibitem{gubser} S. S. Gubser, I. R. Klebanov and A. M. Polyakov,
``Gauge Theory Correlators from Non-Critical String Theory'', Phys. Lett.
{\bf B428} (1998) 105, hep-th/9802109.
\bibitem{our3} P. Minces and V. O. Rivelles, ``Energy and the AdS/CFT
Correspondence'', JHEP {\bf 0112} (2001) 010, hep-th/0110189.
\bibitem{freedman} P. Breitenlohner and D. Z. Freedman, ``Stability in
Gauged Extended Supergravity'', Ann. Phys. {\bf 144} (1982) 249.
\bibitem{freedman8} P. Breitenlohner and D. Z. Freedman, ``Positive Energy
in Anti-de Sitter Backgrounds and Gauged Extended Supergravity'',
Phys. Lett. {\bf B115} (1982) 197.
\bibitem{mezincescu} L. Mezincescu and P. K. Townsend, ``Stability at a
Local Maximum in Higher Dimensional Anti-de Sitter Space and Applications
to Supergravity'', Ann. Phys. {\bf 160} (1985) 406.
\bibitem{witten2} I. R. Klebanov and E. Witten, ``AdS/CFT Correspondence
and Symmetry Breaking'', Nucl. Phys. {\bf B556} (1999) 89, hep-th/9905104.
\bibitem{our2} P. Minces and V. O. Rivelles, ``Scalar Field Theory in the
AdS/CFT Correspondence Revisited'', Nucl. Phys. {\bf B572} (2000) 651,
hep-th/9907079.
\bibitem{gibbons} G. W. Gibbons and S. W. Hawking, ``Action Integrals and
Partition Functions in Quantum Gravity'', Phys. Rev. {\bf D15}
(1977) 2752.
\bibitem{muck} W. M\"uck, ``An improved correspondence formula for AdS/CFT
with multi-trace operators'', hep-th/0201100.
\bibitem{freedman3} D. Z. Freedman, S. D. Mathur, A. Matusis and
L. Rastelli, ``Correlation functions in the $CFT_{d}/AdS_{d+1}$
correspondence'', Nucl. Phys. {\bf B546} (1999) 96, hep-th/9804058.
\bibitem{viswa1} W. M\"uck and K. S. Viswanathan, ``Conformal Field
Theory Correlators from Classical Scalar Field Theory on $AdS_{d+1}$'',
Phys. Rev. {\bf D58} (1998) 041901, hep-th/9804035.
\end{thebibliography}
\end{document}